\documentclass[global,twocolumn]{svjour}

\usepackage{graphics}
\usepackage{amsmath}
\usepackage{epsfig}
\usepackage{amssymb}
\usepackage{dcolumn}
\usepackage{graphicx}
\usepackage{color}

\begin{document}
\title{The key role of off-axis singularities in free-space vortex transmutation}

\author{David Novoa\inst{1,2} \and I\~nigo J. Sola \inst{3} \and Miguel Angel Garc\'{\i}a-March\inst{4} \and Albert Ferrando\inst{5}}                     

\institute{Max Planck Institute for the Science of Light, G\"unther-Scharowsky str. 1, 91058 Erlangen, Germany, \email{david.novoa@mpl.mpg.de} \and
Centro de L\'aseres Pulsados, CLPU. Edificio M3. Calle del Adaja, s/n, 37185 Villamayor, Spain \and
Grupo de Investigaci\'on en \'Optica Extrema (GIOE). Dpto. de F\'{i}sica Aplicada. Univ. de Salamanca. 37008 Salamanca, Spain \and
Dept. d'Estructura i Constituents de la Mat\`eria. Univ. de Barcelona, 08028 Barcelona, Spain \and
Dept. d'\`Optica. Interdisciplinary Modeling Group, InterTech. Univ. de Val\`encia, 46100 Burjassot (Val\`encia), Spain}

\date{Received: date / Revised version: date}

\maketitle
\begin{abstract}
We experimentally demonstrate the generation of off-axis phase singularities in a vortex transmutation process induced by the breaking of rotational symmetry. The process takes place in free space by launching a highly-charged vortex, owning full rotational symmetry, into a linear thin diffractive element presenting discrete rotational symmetry. It is shown that off-axis phase singularities follow straight dark rays bifurcating from the symmetry axis. This phenomenon may provide new routes towards the spatial control of multiple phase singularities for applications in atom trapping and particle manipulation.
\end{abstract}
%
%

\section{Introduction}
\label{intro}
Scalar coherent light beams can present phase singularities, that is, zeros of the complex scalar wave which are recognized by a vanishing intensity around which the phase whirls~\cite{Nye1974,Soskin1997,Soskin2001}. Orbital angular momentum (OAM) can be associated to these wave fields, commonly termed as optical vortices~\cite{Allen1992,Franke-Arnold2008}. The technological applications of optical vortices range from optical tweezers and particle manipulation~\cite{Paterson2001,MacDonald2002,Grier2003} to quantum information~\cite{Mair2001,Molina-Terriza2007}. Moreover, there is a growing interest in using beams with nontrivial phase as building blocks for a new generation of traps for ultracold atoms, where the particles are indeed confined in the dark areas of the beam~\cite{Jaouadi2010,Gaunt2013,Ivanov2013,XHe2013}. Therefore, the external control of the propagation of  phase singularities linked to the regions of the beam where the intensity vanishes is an essential question.

In 3D, the propagation of the phase singularities follow vortex lines or {\it dark rays}. A single optical vortex with $O(2)$ continuous symmetry presents a phase singularity located at the axis of symmetry, and its topological charge corresponds to its OAM. The corresponding dark ray is simply a straight line coinciding with the axis of symmetry. A number of stationary vortices are characterized by these dark rays, which can form knots or braids, whose study is of central interest due to their potential technological applications~\cite{Leach2004,Dennis2010,Padgett2011}. 

These dark rays, considered as the trajectories of vortices propagating through certain medium, can be controlled by means of either non-cylindrically symmetric input wave-fronts or symmetric ones propagating in inhomogeneous or nonlinear media, thus allowing for the inversion of the charge of the initial vortex~\cite{MolinaTerriza2001a,MolinaTerriza2001b,GarciaRipoll2001}.
A different approach considers the controllable transformation of a single optical vortex with $O(2)$ continuous symmetry in a vortex featuring different on-axis charge by means of a system showing discrete rotational symmetry. This phenomenon, known as {\it vortex transmutation}~\cite{Ferrando2005b}, was experimentally demonstrated in optically induced photonic lattices~\cite{Bezryadina2006a,Bezryadina2006} and recently, for the first time, in \emph{free
space} using \emph{linear} polygonal lenses~\cite{Gao2012}. Remarkably, the unveiling of the OAM by diffracting a vortex with a triangular aperture~\cite{Hickmann2010} can be understood in the light of this phenomenon. The process of vortex transmutation, governed by the simple symmetry rule established in~\cite{Ferrando2005b}, was proven 
to be a consequence of the generation of a number of optical vortices on-axis which propagate off-axis ~\cite{Garcia-March2009a}. In this paper we experimentally demonstrate the generation and control of these off-axis singularities (OAS) by means of a \emph{thin} discrete symmetry diffractive element (DSDE) in free space, and show that the corresponding dark rays are straight lines. These results confirm the theory developed in Ref.~\cite{ferrando2013}.


\section{Theory of OAS formation}
\label{sec:1}

\begin{figure}[htb]
\centerline{\includegraphics[width=7.5cm]{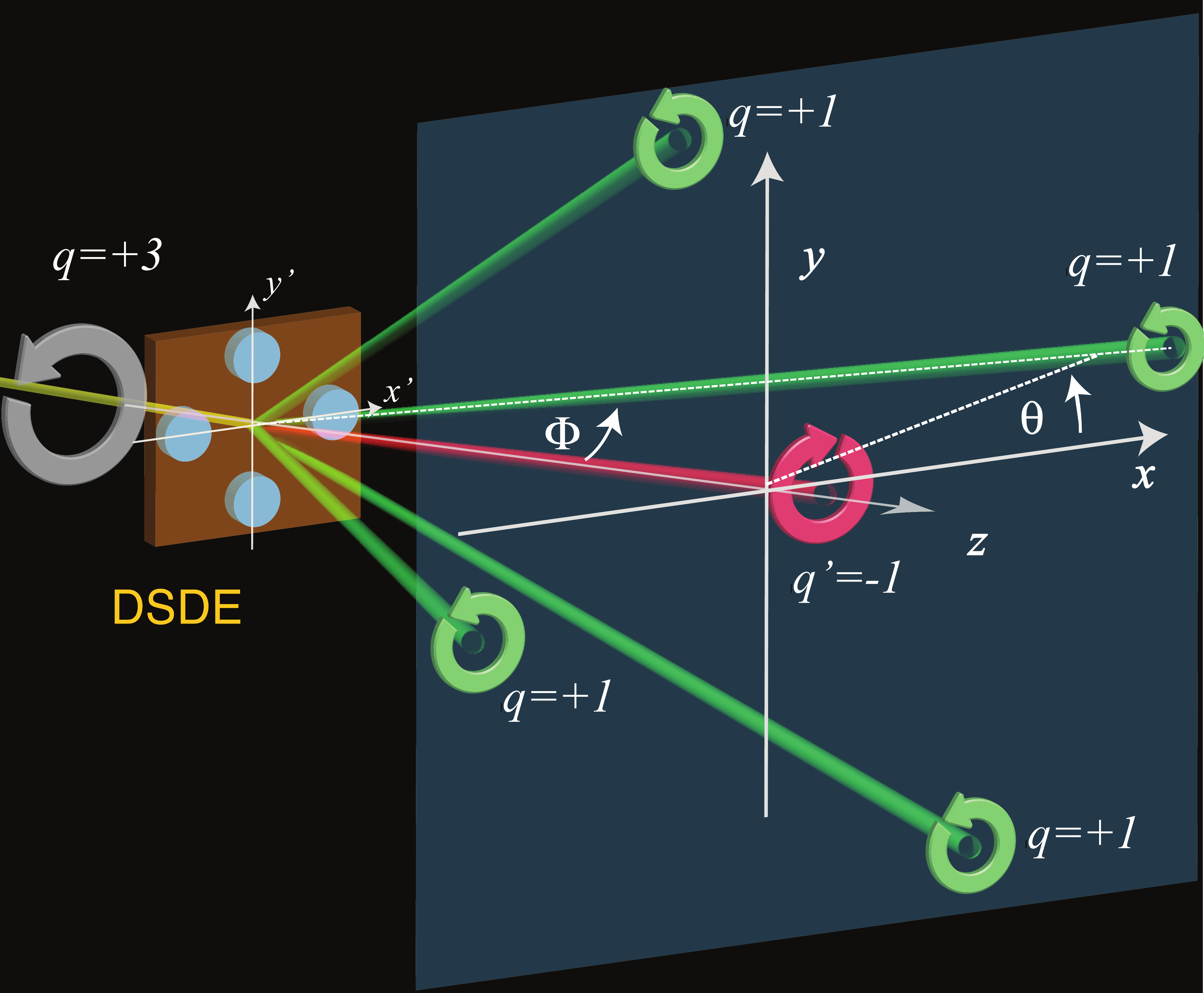}}
\caption{Straight-line trajectories of dark rays after the diffraction of a LG beam with $q=l=+3$ by a thin DSDE with $N=4$. Curved arrows indicate 
the direction of the azimuthal phase variation.}
\label{fig:theory_DSDE}
\end{figure}

The generation of OAS is closely related to the
phenomenon of vortex transmutation \cite{Ferrando2005b}, which occurs
when a highly-charged vortex owning $O(2)$ continuous symmetry illuminates
a medium with discrete rotational symmetry of $N$th order ($\mathcal{C}_{N}$).
The breaking of $O(2)$ into $\mathcal{C}_{N}$ produces four major
effects: (i) OAM is no longer conserved,
and vortex solutions are not eigenfunctions of $L_{z}=-i\partial/\partial\theta$
with eigenvalue $l$ \cite{Garcia-March2009}; (ii) a quantity $m$
called angular pseudo-momentum is conserved instead of $l$ \cite{Ferrando2005c};
(iii) an upper cutoff applies to the topological charge $q$ of a
single phase singularity: $|q|\le N/2$ \cite{Ferrando2005a}; and
(iv), a disintegration of a single phase singularity located at the
symmetry axis occurs when $|q|>N/2$ verifying the following transmutation rule
\begin{equation}
 \label{trans_rule}
 l-m=kN , \quad k\in\mathcal{Z},
\end{equation}

where $q=l$ and $q'=m$ are the charges
of the central singularity before and after the breaking, respectively
\cite{Garcia-March2009a}. Numerical simulations show that
transmutation is accompanied by the appearance of $kN$ OAS
moving outward from the symmetry axis~\cite{Garcia-March2009a}.
Although this result was obtained by assuming illumination of a \emph{thick}
$z$-invariant discrete \emph{nonlinear} medium, the $O(2)\rightarrow\mathcal{C}_{N}$
breaking mechanism is a linear phenomenon. Experimental observation
of the transmutation of the central charge has been recently reported~\cite{Gao2012}. The
theory of OAS in free-space vortex transmutation
has been recently developed in Ref. \cite{ferrando2013}. In this reference, trajectories of the OAS
have been analytically calculated in the case of a \emph{thin} 
DSDE illuminated by a Laguerre-Gaussian (LG) beam with charge $q=l$ \cite{ferrando2013}.
The main results of this theory are summarized in Fig.~\ref{fig:theory_DSDE} for the case $q=+3, N=4$. The dark ray associated to the central singularity does not experience
any deviation after the action of the DSDE. However, $N=4$ ($k=1$) singularities with $q=+1$ follow trajectories bifurcating from the
central ray at the DSDE position. They verify the transmutation rule: $q=q'+kN$ ($q=+3$ and $q'=-1$), conserving charge. Asymptotically, these $N=4$ dark
rays follow straight lines. The polar $\theta$ and axial $\Phi$
angles of the dark rays are completely determined by the parameters
of the DSDE and the input LG beam. In particular, the polar angle of the $j$th dark ray is given by $\theta_{j}=\pi/(2 N)+j\pi/2$
with $j=0,\cdots, N-1$ \cite{ferrando2013}. The axial angle is given by $\Phi=tan^{-1}[\pi^{-1}2\sqrt{6}W_{0}\lambda v^{1/2}]$ ($v\ll1$),
where $W_0$ is the width of the input  LG  beam, $\lambda$ is its wavelength and $v$ is a dimensionless parameter measuring the strength of the $O(2) \rightarrow \mathcal{C}_{N}$ breaking. Remarkably, the dark rays divergence $\Phi\ll1$ is proportional to the waist $W_0$, unlike the beam divergence (bright region) which is proportional to $W_0^{-1}$ (see Ref. \cite{ferrando2013} for details). Thus, dark rays follow an opposite diffraction pattern than their bright counterparts.

\section{Experimental setup}
\label{sec:2}

\begin{figure}[htb]
\centerline{\includegraphics[width=7.5cm]{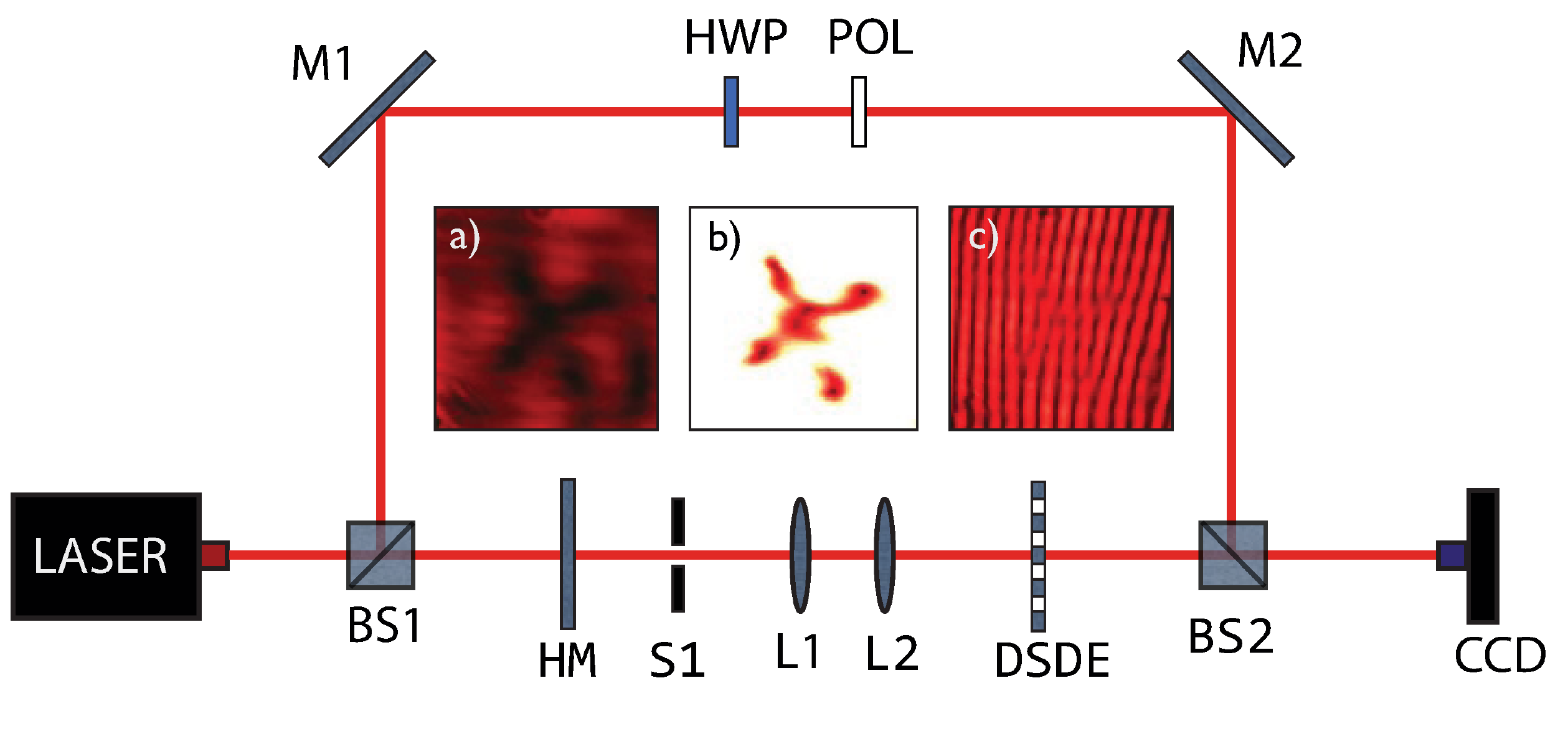}}
\caption{Experimental setup: LASER: 1 mW HeNe laser ($\lambda=632.8$ nm); BS1 and BS2: beam splitters; HM: diffractive holographic
mask; S1: adjustable iris; L1 and L2: lenses with focal lengths f1= 150 mm , f2= 150 mm; 
DSDE: discrete symmetry diffractive element; HWP: half-wave plate; 
M1 and M2: mirrors; POL: linear polarizer; CCD: camera at the observation plane. 
Insets: a) far-field intensity pattern of a vortex with charge $q=+3$ after passing through the DSDE. b) Same as a), except for a higher intensity of the initial vortex beam. c) Interferogram of the situations depicted in a)-b). The fork-like patterns with the spikes up (down) indicate the location of the $q>0$ ($q<0$) phase singularities. All the insets display a $1\times$ 1mm square region.}
\label{exp_setup}
\end{figure}

A sketch of the experimental setup is shown in Fig. \ref{exp_setup}. A HeNe laser ($\lambda=632.8$ nm, $P=1$ mW) was used as a source. The linearly-polarized beam is driven into a Mach-Zenhder interferometer, where a first beam splitter (BS1) divides it into two arms. In one of them, an optical vortex with topological charge $q=+3$ is synthesized by passing the beam through a diffractive mask (HM) \cite{Carpentier2008}. The first diffracted order is selected, blocking the remaining with an aperture (S1). A telescope, made by a couple of lenses L1, L2, allows to control its divergence. In all our experiments, the diameter (divergence) of the initial Gaussian vortex is around $2$ mm  ($1$ mrad). The amplitude DSDE, sketched in Fig. \ref{fig:theory_DSDE} and responsible for the transmutation of the vortex 
according to the selection rules, is a square matrix of black dots (i.e., owning $\mathcal{C}_{4}$ symmetry) 
impressed on a transparent substrate (dot radius=180 $\mu$m; dot-to-dot distance = 880  $\mu$m). 

\section{Generation and dynamics of OAS}
\label{sec:3}

\begin{figure*}[htb]
\centerline{\includegraphics[width=12.5cm]{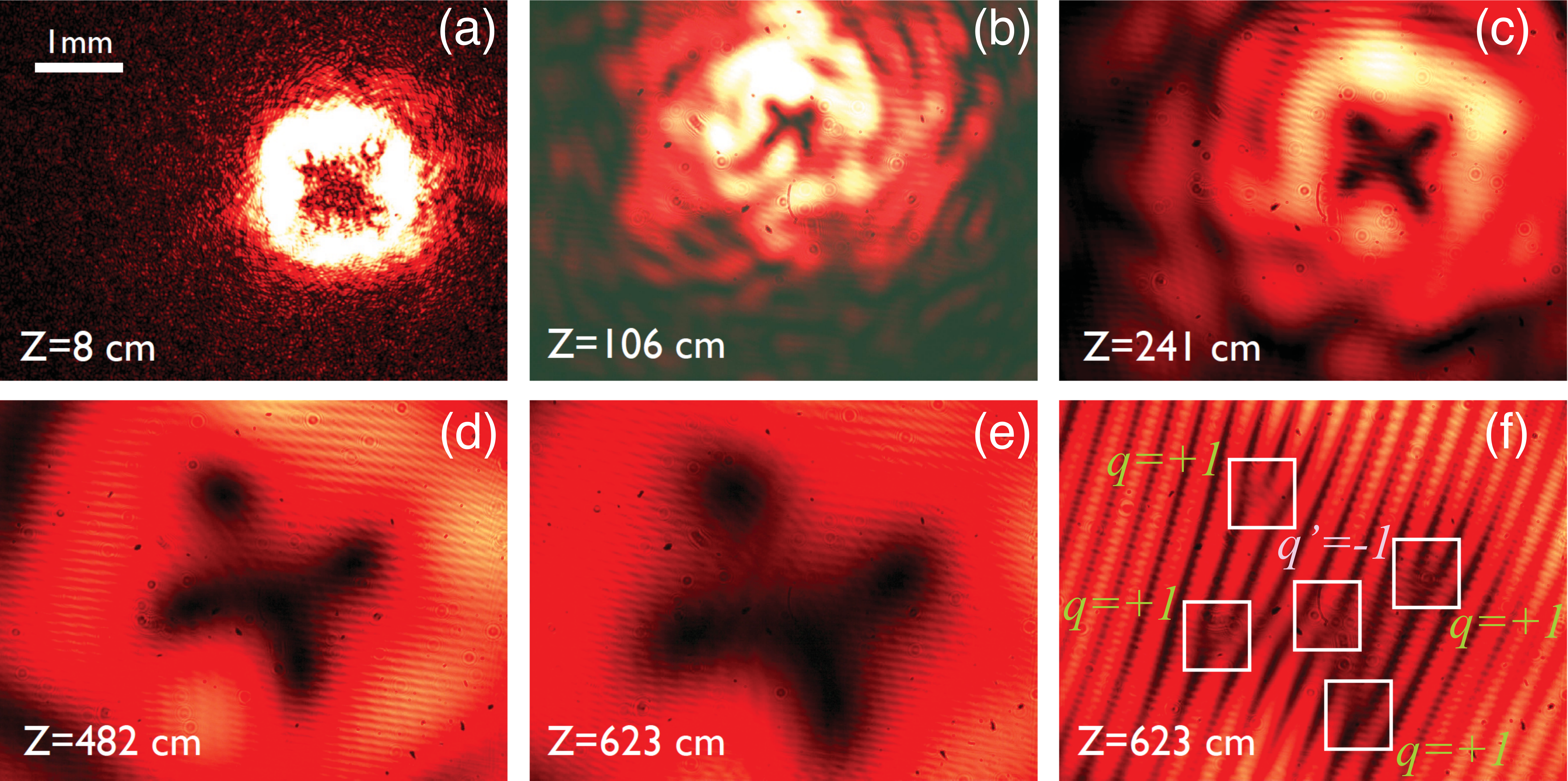}}
\caption{Evolution of phase singularities with propagation: (a)-(e), far-field intensity patterns for increasing values of $z$; (f) interferogram with a plane wave unveiling the value of individual charges. }
\label{fig:vortex_diffraction_images}
\end{figure*}

In the insets of Fig. \ref{exp_setup}, we display the transverse intensity 
distribution recorded by the CCD at a distance $z=75$ cm from the beam splitter BS2. The output pattern is composed by five regions where intensity vanishes 
(see Fig. \ref{exp_setup}a), which, in addition, turn out to be the locations of the arising phase singularities. Because the DSDE employed in our experiments is an amplitude object, the obtained far-field diffraction patterns are strongly distorted, and so additional diagnostics are needed in order to ensure the resulting phase distribution. Notice that a clear analysis of the latter is crucial to have a detailed insight of the vortex transmutation process. Thus, we have adopted two complementary strategies to unambiguously discriminate the phase singularities from other low-intensity regions in the observation plane. First, as singularities are zeros of the optical field, we can identify them clearly by increasing the beam intensity, thus improving the background contrast accordingly (see Fig. \ref{exp_setup}b). Second, we were able to observe directly both positions and charges of the phase singularities using the interferometric setup displayed in Fig. \ref{exp_setup}. Hence, the vortices are identified 
via the fork-like patterns they produce as a result of their interference with a tilted light beam without nested singularities, coming from the second arm of the interferometer. As predicted by theory (see Fig.~\ref{fig:theory_DSDE}), the central dislocation transmutes its charge from $q=+3$ to $q'=-1$, and $N=4$ OAS with charge $q=+1$ each bifurcate from the original vortex $q=+3$ at the position of the DSDE according to its $\mathcal{C}_{4}$ symmetry (see Fig. 2c). Our procedure delivers a fully reliable picture of the whole phase portrait since the interferometric resolution can be tuned at each observation distance by adjusting the relative angle between the beams coming from both interferometer arms.

\begin{figure}[htb]
\centerline{\includegraphics[width=7.5cm]{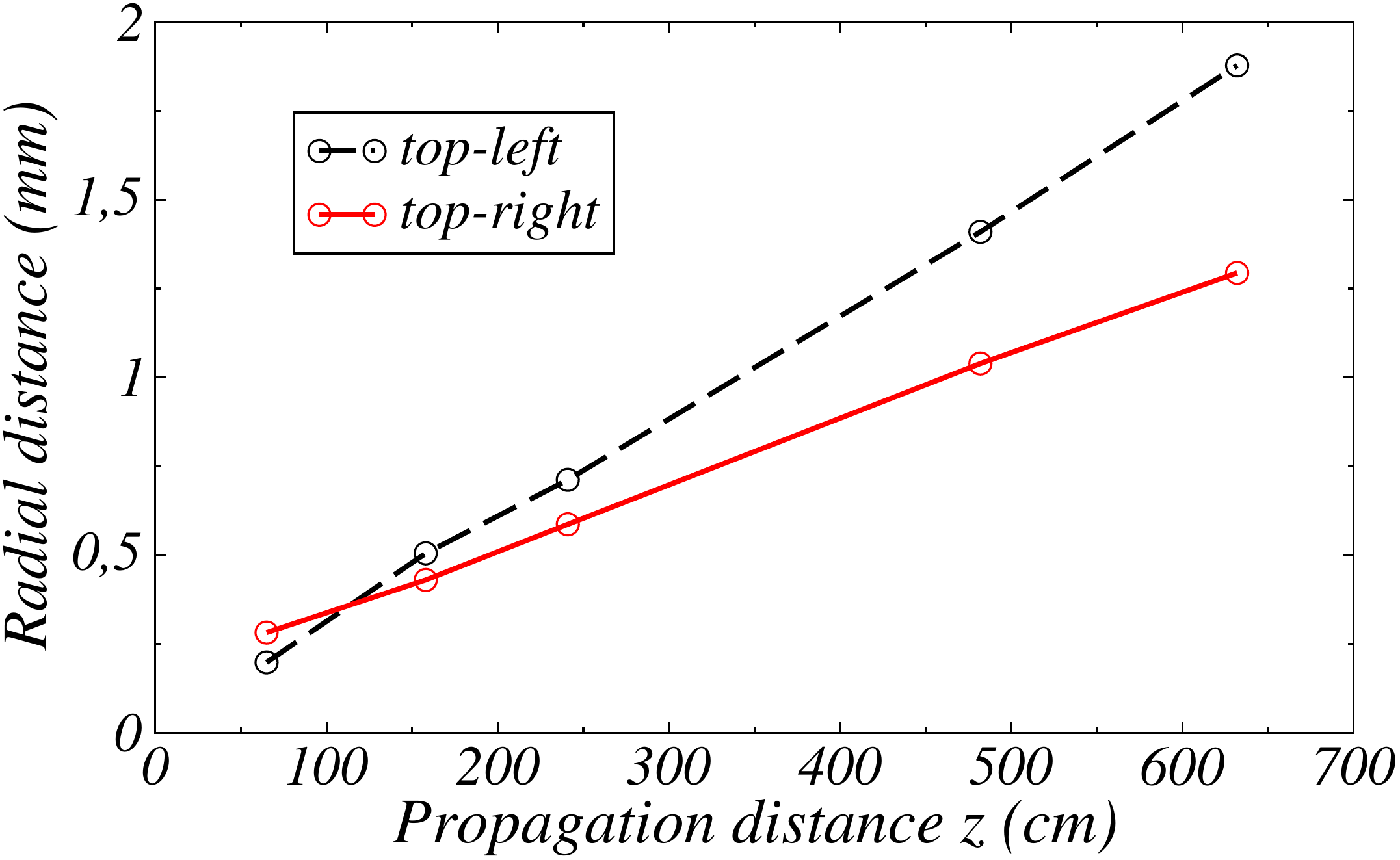}}
\caption{Linear dependence of the radial position on the propagation distance for the top-left and top-right singularities in Fig. ~\ref{fig:vortex_diffraction_images}. The bottom singularities, whose evolution is omitted for simplicity, follow the same linear behavior with slightly different slopes.}
\label{fig:off_singularities_position}
\end{figure}

We have also analyzed the dynamics of the OAS by following their evolution over long distances. The results are depicted in Fig. \ref{fig:vortex_diffraction_images}. Remarkably, the central transmuted singularity does not change the initial trajectory remaining at the origin, while OAS do deviate (see interferogram in Fig. \ref{fig:vortex_diffraction_images}(f)). As experimentally demonstrated in Fig. 4, OAS follow straight dark rays preserving
 $\mathcal{C}_{4}$ symmetry. Particularly,
they display a different behavior of their divergence with respect to the waist $W_0$ when 
compared to the diffraction of the bright ring of the beam (see  Sec.~\ref{sec:1}).
 Differences in the slopes of the lines represented in Fig. \ref{fig:off_singularities_position} can be attributed to distortions caused by the high sensitivity to alignment of the setup.

We must point out that  the solely monitoring of the phase singularity placed upon the optical axis is not enough to demonstrate the transmutation process,
since the transformations in this singularity are compatible with the decay of the initial highly-charged vortex due to a small perturbation~\cite{Freund1999}. In fact, we checked that a small misalignment of the setup causes an immediate conversion of the initial charge $q=+3$ into three single charges $q=+1$. Such behavior turns out to be very similar to that reported in Refs.~\cite{Bekshaev2004,Bekshaev2008}, where the authors observed the decay of a single vortex of charge $|m|>1$ into $|m|$ single-charged vortices due to astigmatism. Thus, some remnant aberrations of our telescopic system, together with the symmetry breakdown from the misaligned DSDE, may explain the decay to $3$ single-charged vortices of our initial vortex beam. Then, the vortex transmutation phenomenon discussed in this paper clearly overcomes that astigmatic breaking mechanism when the system is properly aligned and almost free of major optical aberrations. In such case, the observation of OAS allows to disentangle both phenomena 
without any ambiguity.

Finally, we observed a slight rotation of the whole pattern (see Fig. \ref{fig:vortex_diffraction_images}). Whilst such rotation can be compatible with the theoretical description, some distortion due to the optical system can also yield to similar results.

\section{Conclusions}
\label{conclus}
We have experimentally demonstrated the generation of OAS in free-space vortex transmutation by means of a DSDE. We have shown that OAS follow straight dark rays in the form predicted by recent theory ~\cite{ferrando2013}. Our study opens the door to the full control over both the generation and evolution of deterministic patterns of phase singularities by means of DSDEs. These results can lead to relevant applications in fields such as particle manipulation, quantum information, and ultracold atom trapping, in which the non-trivial phase pattern of new types of optical beams characterized by multiple phase singularities, as those presented here, can play a key role.


\begin{acknowledgement}
A. F. and M.A.G.M. acknowledge support from the Spanish grants TEC2010-15327 and FIS2011-24154. M.A.G.M. also acknowledges grant No. 2009SGR-1289 from Generalitat de Catalunya. D. N. acknowledges support from MINECO through FCCI. ACI-PROMOCIONA (ACI2009-1008). I. J. S. acknowledges MINECO for Consolider Program SAUUL (CSD2007-00013), project FIS2009-09522 and Ram\'on y Cajal grant. We thank H. Michinel and A. V. Carpentier for providing us with the holographic masks.
\end{acknowledgement}


\end{document}